\documentclass[12pt]{iopart}
\usepackage{iopams}  
\usepackage{graphicx}

\graphicspath{ {./figures/} }

\begin{document}

\title[Stearyl--Cetyl Alcohol--Polysorbate 60--Water Ternary System]
{
Influence of fatty alcohol mixing ratios on physicochemical properties of 
Stearyl-Cetyl-Polysorbate 60-Water ternary System: 
Insights from Experiments and Computer Simulations
}

\author{Vu Dang Hoang$^{1*}$, Cao Phuong Cong$^2$, Hung Huu Tran$^3$, Hue Minh Thi Nguyen$^3$, 
and Toan T. Nguyen$^{2*}$}
\address{$^1$ Department of Analytical Chemistry and Toxicology, Hanoi University of Pharmacy, 13-15 Le Thanh Tong street, Hanoi, Vietnam}
\ead{hoangvd@hup.edu.vn}
\address{$^2$ Key Laboratory for Multiscale Simulation of Complex Systems, VNU University of Science, Vietnam National University, 334 Nguyen Trai street, Thanh Xuan, Hanoi 120401, Vietnam}
\eads{\mailto{toannt@hus.edu.vn},\mailto{toannt@vnu.edu.vn}}
\address{$^3$ Faculty of Chemistry and Center for Computational Science, Hanoi National University of Education, 136 Xuan Thuy street, Cau Giay, Hanoi, Vietnam}

\begin{abstract}
The structure and stability of ternary systems prepared with polysorbate 60 and various combinations of cetyl (C16) and stearyl (C18) alcohols were examined as they aged over 3 months at 25$^\circ$C. 
Rheological results showed that the consistency of these systems increased initially during roughly the first week of aging, which was succeeded by little changes in consistency (systems containing from 30\% to 70\% C18, 
with the 50\% C18 system showing the highest consistencies)
 or significant breakdown of structure (remaining systems).
The formation and/or disintegration of all ternary systems were also detected by microscopy 
and differential scanning calorimetry experiments. 
This study emphasizes the fact that the structure and consistency of ternary systems 
are dominantly controlled by the swelling capacity of the lamellar gel phase. 
Molecular dynamics simulations were performed to provide details on the molecular mechanism of stability.
Computational results support the hypothesis experimentally proposed 
for the stability of the mixed system being due to an increase flexibility of the alcohol 
with a longer hydrocarbon chain.
 Computational results are in excellent agreement with experimental conclusions. 
 Additionally, they show that in the mixed system, 
 alcohols with shorter hydrocarbon chain become more rigid. 
 These molecular details could not be available in experimental measurements.
\end{abstract}

%
\vspace{2pc}

%
%
%

\section{Introduction}\label{intro}

Ternary systems containing surfactants and fatty alcohols in excess water are widely used in pharmacy and cosmetics
(e.g. skin and hair care products) as a means of topically and transdermally delivering an active ingredient directly 
to the skin \cite{Engels1998}.
As a result, their physicochemical properties, rheology, and multiphase structure have attracted strong research interests 
recently \cite{Ahmadi2020,Datta2020,Iwata2017,HoangVu2020}.
These systems, generally speaking, could be employed as models to represent the continuous phases of the respective semisolid o/w emulsions. 
Optimizing the ratios of their various components to attain a physically stable o/w emulsion is one of the main research directions of 
these systems \cite{Korhonen2000}.

In industry, one can see the changes in consistency of topical hydrophilic creams prepared with different batches of excipients. 
It does mean that excipient batch-variation may not only affect the self-life of creams but also probably vary their bioavailability. 
Previous publications on characterizing oil-in-water emulsions formulated from mixed and pure homologue fatty alcohols 
(i.e. stearyl C18, cetyl C16, cetostearyl CSA) and ionic surfactants or non-ionic surfactants
\cite{Barry1970, Eccleston1976, Eccleston1982, Eccleston1985, Fukushima1977, Eccleston1977, Fukushima1976, Sepulveda2003, Talman1970}
have demonstrated that the consistency of these topical formulations established as semisolids and remained nearly unchanged on the use of CSA; 
whereas they changed substantially from semisolids (on the preparation) to mobile liquids (on further storage) when C16 or C18 was individually used.
Even though the influence of various fatty alcohol mixing ratios on the structure and stability of oil-in-water emulsions 
was partly shown in the above-mentioned works, further investigation 
is still required when different surfactants are employed in cream formulae.
 
In theory, the o/w type creams stabilized with emulsifying waxes and mixed emulsifiers consisted of surfactants and fatty alcohols have no less than four phases: (i) hydrophilic/crystalline gel phase composed of bilayers of fatty amphiphile and surfactant molecules that incorporates water molecules to form interlamellar water; (ii) free-bulk water pockets in balance with the interlamellarly fixed water; (iii) lipophilic gel phase composed of the excess of the fatty amphiphile; and (iv) dispersed oil phase as droplets surrounded by multilayers of gel phase \cite{Eccleston1997}.
To evaluate the physical stability and microstructure of topical creams, commonly used techniques are rheology, 
differential scanning calorimetry (DSC), and microscopy \cite{Ballman2008, Estanqueiro2014, Masmoudi2006, Konya2003, Olejnik2017, Olejnik2015, Ribeiro2004, Savic2005, Savic2009}.

In practice, rheological assessment is a conventional means of characterizing pharmaceutical semisolids. 
It provides an indirect understanding of the structure and parametric specification on the consistency, 
influencing sensory attributes and drug release. 
For a pharmaceutical gel, mechanical and viscoelastic properties can be rheologically characterized by flow curve (shear stress vs. shear rate), 
creep and recovery curve, and oscillation measurements 
[the elastic response (storage modulus, $G'$) and viscous behavior (loss modulus, $G''$) in the linear viscoelastic region]. 
The evidence about the liquid-crystalline or crystalline state of this system can be accessed 
by DSC measurements on transition temperatures and also enthalpy changes at these transitions. 
In addition, its anisotropic structure (e.g. mosaic texture, oily streaks, and Maltese crosses) that implicates the presence of lamellar phase, 
could be well revealed by polarised light microscopy. 

Computational pharmaceutics, especially computer modeling and simulation using atomistic models, 
are important tools to provide additional insights at molecular level to experimental data \cite{Ouyang2015}. 
Molecular dynamics simulations can elucidate not only the molecular mechanism of drug delivery systems 
but also the atomic structures and interactions of these systems, which is very useful 
in designing experiment formulations as well as simplifying formulation screening procedures \cite{Zhao2018}.

The objective of this study was to investigate the influence of fatty alcohol mixing ratios 
on the physicochemical properties of stearyl--cetyl alcohol--polysorbate 60--water ternary systems. 
They were experimentally investigated by using macro- and microscopy, rheology, and
differential scanning calorimetry as being aged over 3 months at ambient temperature. 
Then, to better understand the underlying mechanism for the stability of our ternary system, 
computational molecular dynamics investigation was carried out using atomistic model in explicit solvent. 
Various parameters of ordering in the mixed systems namely the deuterium order parameter, area per lipid, 
magnitude of fluctuations are computed. The fact that the mixed systems show higher stability 
than the pure systems could be explained by a mismatch in the alkyl chain length as stated in the current experimental hypothesis \cite{Iwata2017, Snyder1983}.
Consequently, a high degree of freedom should be seen for the alkyl chain tip of C18OH, 
hence higher configurational entropy. 
In our study, computer simulation data support this hypothesis and also reveal 
many additional molecular details such as the higher fluctuations of the bilayer molecules in mixed systems. 
The shorter alkyl chain becomes more rigid upon mixing, which actually helps further increase configurational space for the longer chain. These details could not be available from experiments. Both of our extensive experiments and computational investigations of ternary systems at various mixing ratios of C16 and C18 fatty alcohol show that the 50:50 mixing ratio gives the optimal formulation with the highest consistency. 

Combining experiments and computer simulation provide a comprehensive, complementary understanding of these systems at macroscopic and molecular levels, is of the first of its kind for these ternary systems to our knowledge. 

The paper is organized as follows. After the introduction section, in section \ref{prepare}, the model of our system and various physical parameters used in the simulation are presented in detail. In section \ref{result}, the results of various experimental measurements and computer simulation are presented. In section \ref{discussion}, the results are discussed in detail, providing complementary theoretical and experimental pictures of the underlying mechanism. We conclude in section \ref{conclusion}.

\section{Materials and methods}\label{prepare}

\subsection{Materials}
Fatty alcohols (cetyl, C16 (99\%) and stearyl, C18 (99\%)) and polysorbate 60 of pharmaceutical-grade were kindly provided from Croda, Singapore. Freshly distilled water was used throughout. 

\subsection{Preparation of fatty alcohol admixtures}
In this study, fatty alcohol admixtures were prepared by mixing the molten mixtures of C18:C16 combinations (w/w \%) (i.e. 10:90, 30:70, 50:50, 70:30, 80:20, 85:15, and 90:10) and leaving them to solidify at room temperature. To make a representative hydrated fatty alcohol admixture (50:50 C18:C16), an equal amount of water was added to the corresponding amount of a molten mixture of fatty alcohols.

\subsection{Preparation of ternary systems}
Fatty alcohol (16g) -- polysorbate 60 (4g) -- water (180g) ternary systems were prepared by adding a 70\textsuperscript{o}C molten mixture of fatty alcohol and Polysorbate 60 to hot water and homogenizing with the use of a Silverson SL2T (Silverson machines, Chesham, UK) set at 1000rpm until cooling down to room temperature. During homogenization, no compensation was made for the evaporative water loss (being considered minimal). 
To investigate the influence of C18-C16 combinations, ternary systems were prepared with polysorbate 60 and various combinations of C18:C16 (w/w \%) i.e. 0:100, 10:90, 30:70, 50:50, 70:30, 80:20, 85:15, 90:10 and 100:0.

\subsection{Macroscopic and microscopic observation}
The ternary systems were examined visually and microscopically using polarized and normal light with a Polyvar system (Reichart--Jung, Austria) throughout the storage time. 

\subsection{Rheological tests}
Rheological tests were performed at 25\textsuperscript{o}C using a CSL 100 Rheometer, TA Instrument, UK with 4cm, 2\textsuperscript{o} cone and plate. 
A DHR solvent trap was used throughout the tests to minimize water evaporation.

\subsubsection{Continuous shear}
Flow curves (shear rate vs. shear stress) were recorded when the samples were subjected to an increase and a decrease in shear rate sweep in the range 0.3--100s\textsuperscript{-1} in 10 minutes (5 minutes up; 5 minutes down).

\subsubsection{Oscillation}
To determine the viscoelastic region, screening tests were done at selective frequencies 0.1 and 10Hz with strain varied from 10\textsuperscript{-5} to 10\textsuperscript{-1}rad. Unless stated otherwise, samples were tested under oscillation frequency sweep mode (i.e. 0.01--10Hz) at 2.5×10\textsuperscript{-5}rad to obtain oscillation data such as $G'$ and $\eta'$.

\subsection{Differential scanning calorimetry (DSC) analysis}
Samples of ca. 5mg fatty alcohols (i.e. pure C16, C18; C18:C16 admixtures; hydrated 50:50 C18:C16 admixture) and 10mg ternary systems were placed in hermetically sealed pans. DSC experiments between 15--80\textsuperscript{o}C were performed with 3 cycles: heating, cooling, and re-heating at the rate of 10\textsuperscript{o}C/min using a TC 15 (Mettler Toledo, Switzerland).

\subsection{Molecular dynamics simulation}
For C16OH and C18OH modeling, their chemical structures are obtained from the PubChem database (https://pubchem.ncbi.nlm.nih.gov/) with CID 2682 and 8221, respectively. For polysorbate 60 molecule, its chemical structure is built manually. These three-dimensional chemical structures are then optimized and added partial charges of atoms using Gaussian 16 software package \cite{g16}. For Gaussian computation, the basis set 6-31G and the B3LYP functionals are used \cite{Rassolov2001, Stephens1993}. The Restrained Electrostatic Potential (RESP) is used for the calculation of the partial charge for its high compatibility with the AMBER force fields that we employed \cite{Woods2000}. The obtained partial charges are then manually rounded to make sure that they are the same for every hydrocarbon tails of the alcohols and polysorbate 60 because of their similarity in chemical structure and function. They also match the Slipids force field for lipids and with the united atom model for polyalcohols because these force fields are widely accepted and compatible with the AMBER force fields \cite{Ferrando2009, Jambeck2012}. Except for the hydrocarbon tail, the partial charges of the polysorbate 60 molecule are kept unchanged as obtained from the quantum mechanical calculation. For the bonded interactions, the Amber GAFF force fields from AmberTools 2017. The common TIP3P force field is used to parameterize water molecules due to its compatibility with the AMBER force fields.

The initial configuration of a bilayer made from C16, C18, and polysorbate 60 is constructed manually as follows. A hexagonal lattice is prepared to present the position of each molecule in the lattice. The lattice is in the xy-plane with a lattice constant of 0.42nm. The lattice size of the bilayer is $14\times14$ considering possible computational cost. The hydrocarbon tails are hydrophobic so that they head to the inside of the bilayer along the z-axis direction, whereas the hydroxyl head groups are hydrophilic heading to the outside of the bilayer. A number of systems with different ratios of C18 and C16 are prepared, namely 10:90, 30:70, 50:50, 70:30, and 90:10. The polysorbate 60 molecule plays a role in making the bilayer stable. The number of polysorbate 60 molecules takes about 5\% of the total number of molecules of a bilayer (20\% of the total mass of a bilayer) to match the experimental systems. Due to the large size and complex hydrophilic tails of polysorbate 60, the initial configuration of a bilayer is carefully prepared by rotating and translating the polysorbate 60 molecules out of the bilayer within a reasonable range of 1.5nm to prevent the overlapping among the hydrophilic branches. The bilayer is then solvated with water, which takes up to 90\% of the total mass of the system according to the experimental systems.

Periodic boundary condition is set up for all three directions of the systems. The thickness of water is equal to about 28nm along the z-axis, in the range of typical experiment systems from 6--100nm \cite{Eccleston2000}. Therefore, the electrostatic coupling between the bilayers in neighboring simulation boxes is minimal. The number of each type of molecule of each mixture ratio is described in detail in \tref{table2}. For long-range electrostatic interaction, the Particle Mesh Ewald method is used with the cut-off range of 1.2nm \cite{Darden1992}. The cut-off range of van der Waals interactions is also 1.2nm. All covalent bonds with hydrogen atoms are constrained during simulation using the LINCS algorithm \cite{Hess1997}.

\Table{\label{table2} The ratio of molecules in a $14\times14$ membrane. The number of molecules are calculated from the ratio of molecular masses from the experiments. }
\br
\multicolumn{1}{l}{\begin{tabular}[l]{@{}l@{}}C18\\(\%)\end{tabular}} &
\multicolumn{1}{l}{\begin{tabular}[l]{@{}l@{}}C16\\(\%)\end{tabular}} &
\multicolumn{1}{l}{\begin{tabular}[l]{@{}l@{}}C18\\(molecule)\end{tabular}} &
\multicolumn{1}{l}{\begin{tabular}[l]{@{}l@{}}C16\\(molecule)\end{tabular}} &
\multicolumn{1}{l}{\begin{tabular}[l]{@{}l@{}}Polysorbate 60\\ (molecule)\end{tabular}} &
\multicolumn{1}{l}{\begin{tabular}[l]{@{}l@{}}Membrane\\ (molecule)\end{tabular}} &
\multicolumn{1}{l}{\begin{tabular}[l]{@{}l@{}}Water\\ (molecule)\end{tabular}} \\ \mr
\enspace 10 & 90 & 34  & 340 & 18 & 392 & 33215 \\ 
\enspace 30 & 70 & 104 & 270 & 18 & 392 & 33124 \\ 
\enspace 50 & 50 & 176 & 198 & 18 & 392 & 33031 \\ 
\enspace 70 & 30 & 252 & 122 & 18 & 392 & 32978 \\ 
\enspace 90 & 10 & 332 & 40  & 20 & 392 & 32747 \\ \br
\endTable

For the simulations, the GROMACS 2018.3 molecular dynamics software package is used \cite{Abraham2015, Berendsen1995}. Starting from the initial configurations, the systems are geometrically optimized, then equilibrated at temperature 300K and pressure 1atm using NPT ensemble. Because of the very far-from-equilibrium configuration of the system, in the beginning, the time-step is conveniently adjusted to gradually increase from 0.2fs to 2.0fs. The total amount of time for this equilibration is about 100ns. The Berendsen thermostat and barostat are used isometrically along all three dimensions. It means that the bilayer can shrink or stretch independently along the x-axis, y-axis, and z-axis of the system allowing the system to relax into its optimal structure. For the production run, the Nose-Hoover thermostat and Parrinello--Rahman barostat are applied. The temperature and pressure are kept the same as the equilibrating procedure.

\section{Results}\label{result}
\subsection{Macroscopic and microscopic observation}
\begin{figure}[!ht]
	
	\centering
	\begin{tabular}[b]{c}
		\includegraphics[width=.45\linewidth]{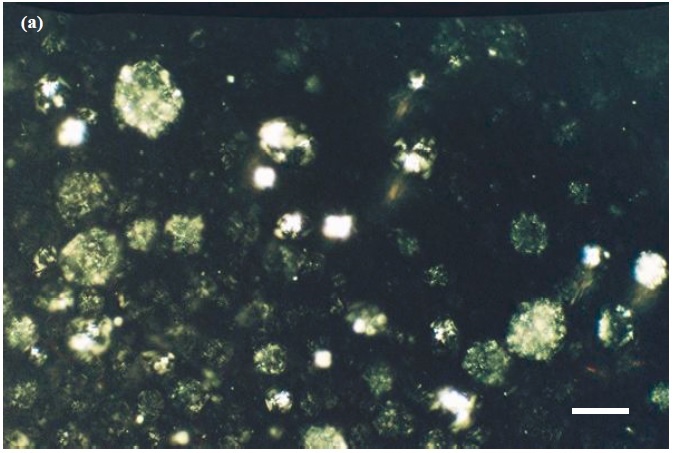} \\[\abovecaptionskip]
		\small (a)
	\end{tabular}
	\begin{tabular}[b]{c}
		\includegraphics[width=.45\linewidth]{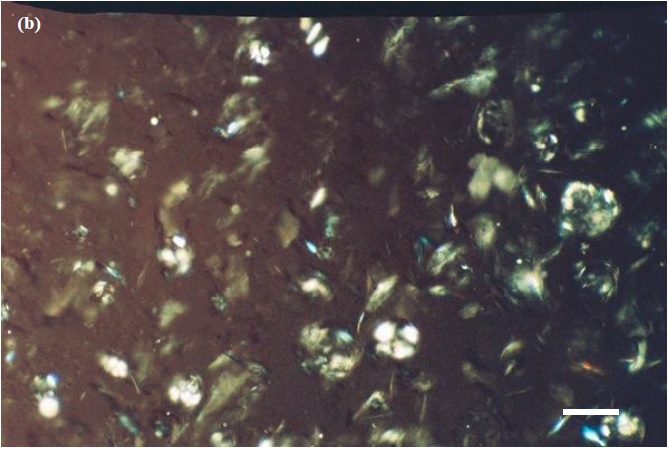} \\[\abovecaptionskip]
		\small (b)
	\end{tabular}
	
	\caption{Typical photomicrographs of a stable system 50:50 C18-C16 (figure a) and an unstable system 10:90 C18-C16 (figure b). Bar: 50$\mu m$.}
	\label{fig1}
\end{figure}

All ternary systems were shiny white semisolid immediately after preparation. Little change was visually observed after 3-month storage for systems containing 30--70\% C18 except that their appearance was slightly glossy with age. In contrast, the others gradually became pearlescent and less solid (10\%, 80\% C18 after 1-month storage), or rapidly changed to mobile systems ($\geq$ 85\% C18 and 100\% C16 systems). In all ternary systems, microscopy revealed scattered masses of spherical structures that were Maltese crosses in polarized light. In addition, large anisotropic "clumps" were visible; some of which were masses of Maltese crosses and others were less well defined (\fref{fig1}a). For unstable systems (10\%, $\geq$ 80\% C18 and 100\% C16), some crystallization appeared as tiny needles or diamond-shaped crystals (\fref{fig1}b) on storage as the systems broke down.

\subsection{Rheological data}
\begin{figure}[!ht]
	\centering
	\begin{tabular}[b]{c}
		\includegraphics[width=.45\linewidth]{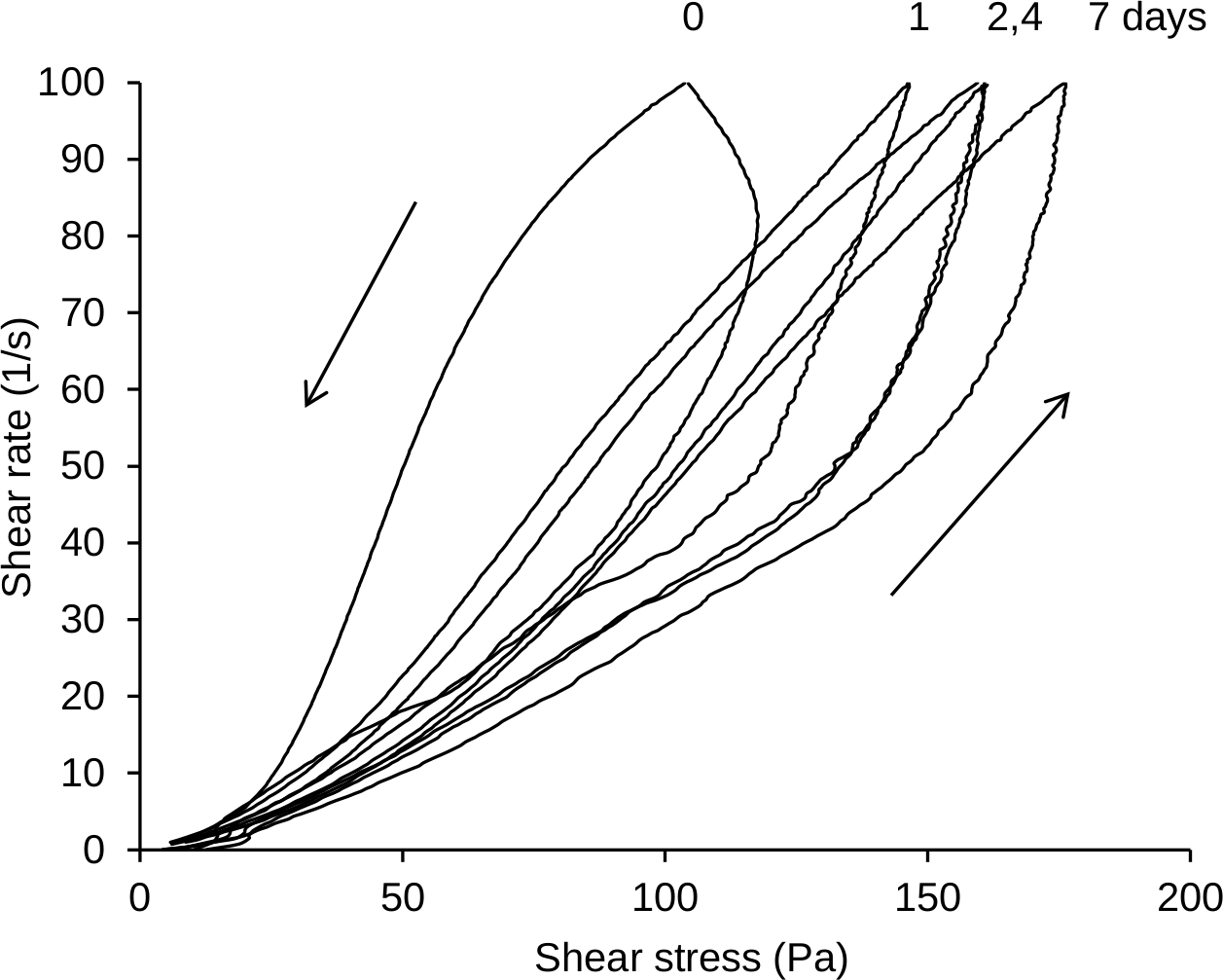} \\[\abovecaptionskip]
		\small (a)
	\end{tabular}
	\begin{tabular}[b]{c}
		\includegraphics[width=.45\linewidth]{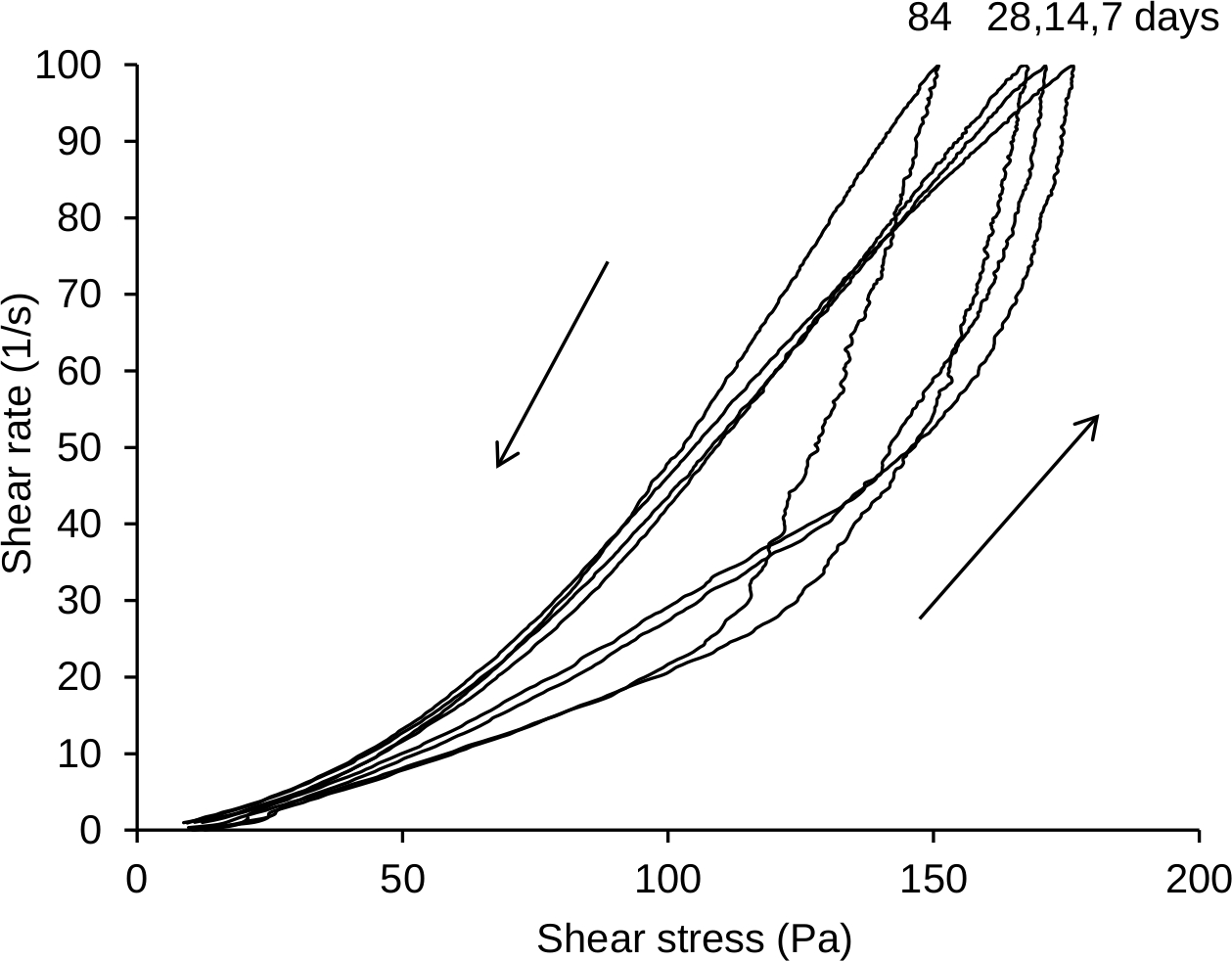} \\[\abovecaptionskip]
		\small (b)
	\end{tabular}
	\caption{Continuous shear tests of 50:50 C18-C16 ternary systems over 3-month storage.}
	\label{fig2}
\end{figure}

\begin{figure}[!ht]
\centering
\begin{tabular}[b]{c}
	\includegraphics[width=.45\linewidth]{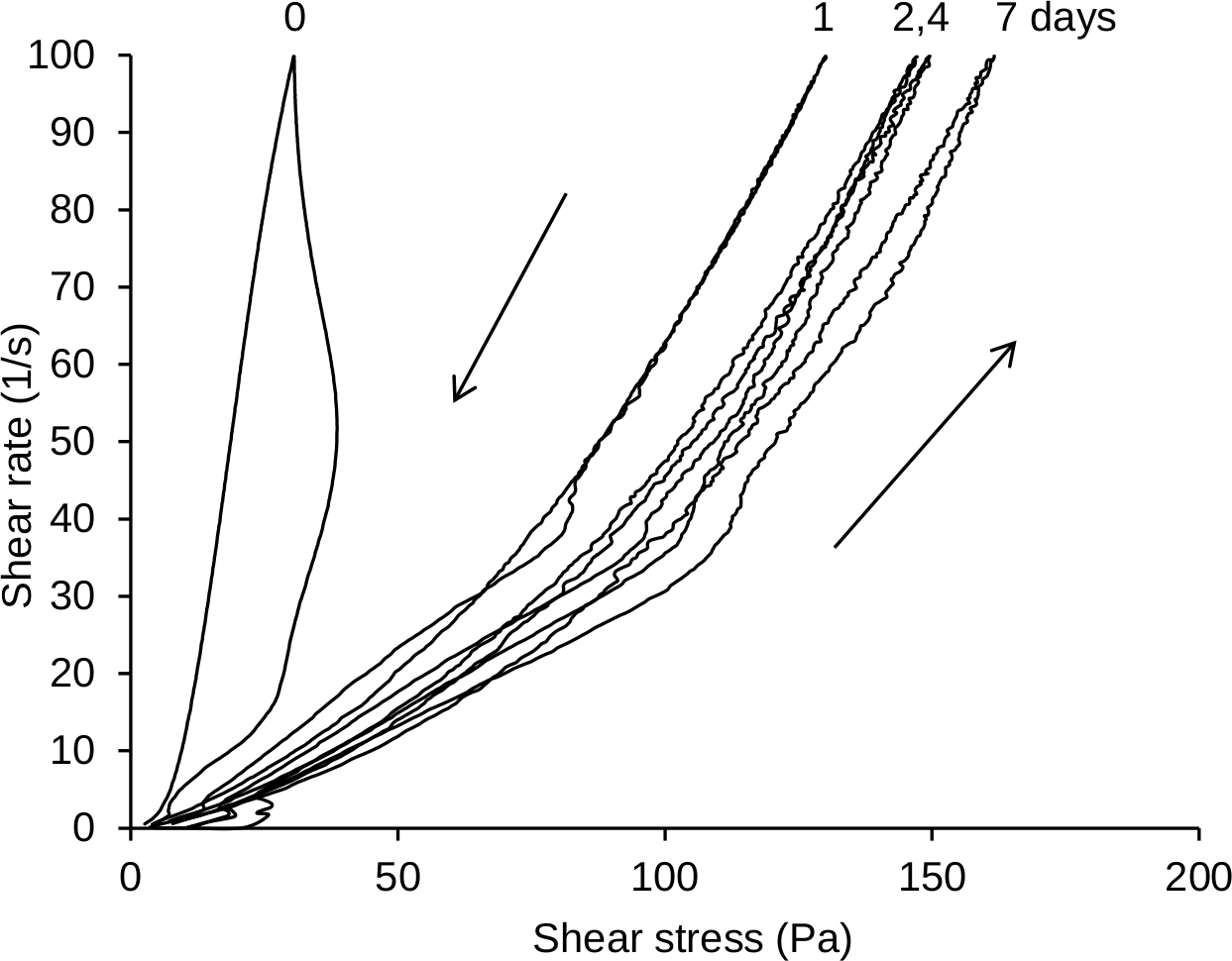} \\[\abovecaptionskip]
	\small (a)
\end{tabular}
\begin{tabular}[b]{c}
	\includegraphics[width=.45\linewidth]{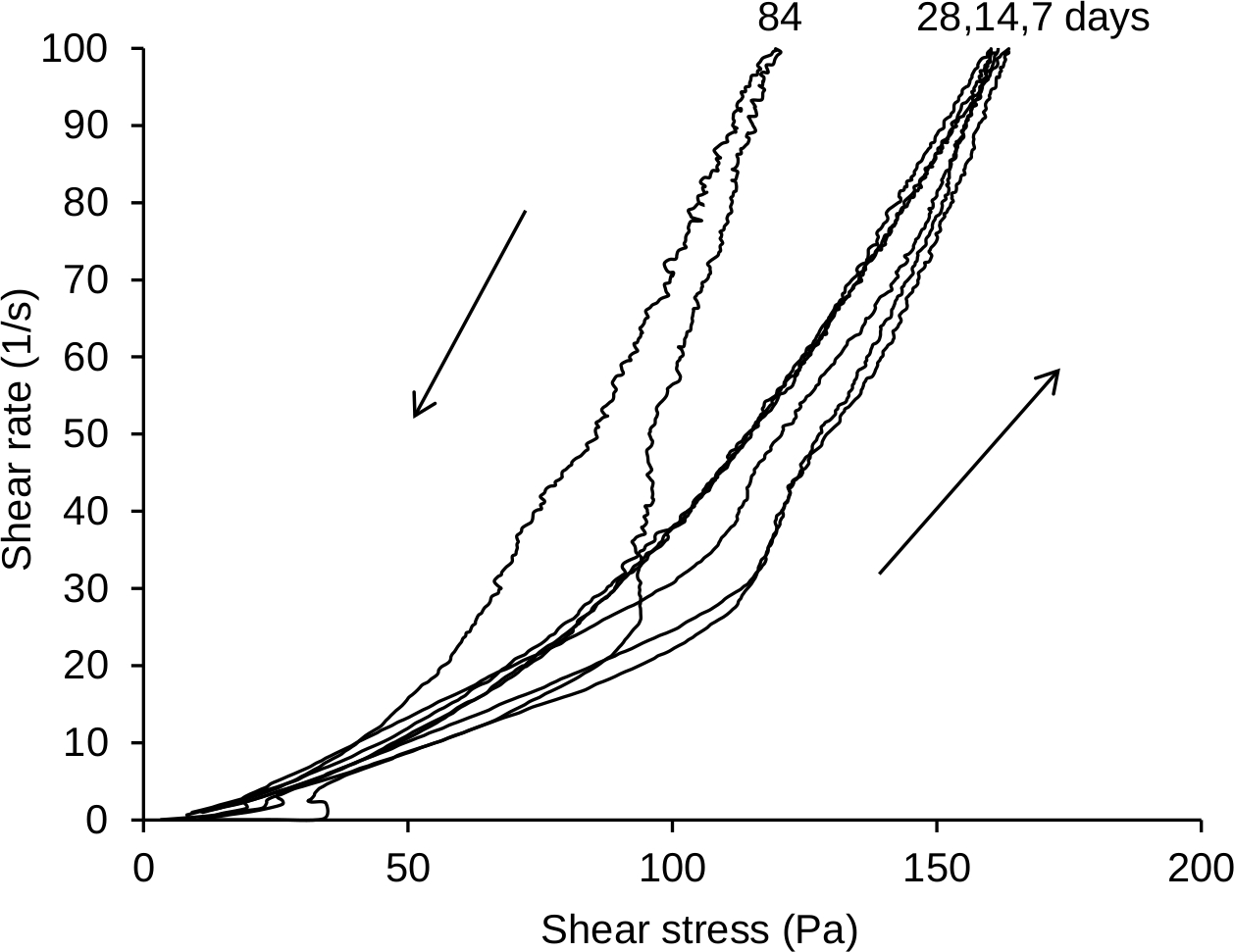} \\[\abovecaptionskip]
	\small (b)
\end{tabular}
\caption{Continuous shear tests of 10:90 C18-C16 ternary systems over 3-month storage.}
\label{fig3}
\end{figure}

\begin{figure}[!ht]
	\centering
	\includegraphics[width=.8\linewidth]{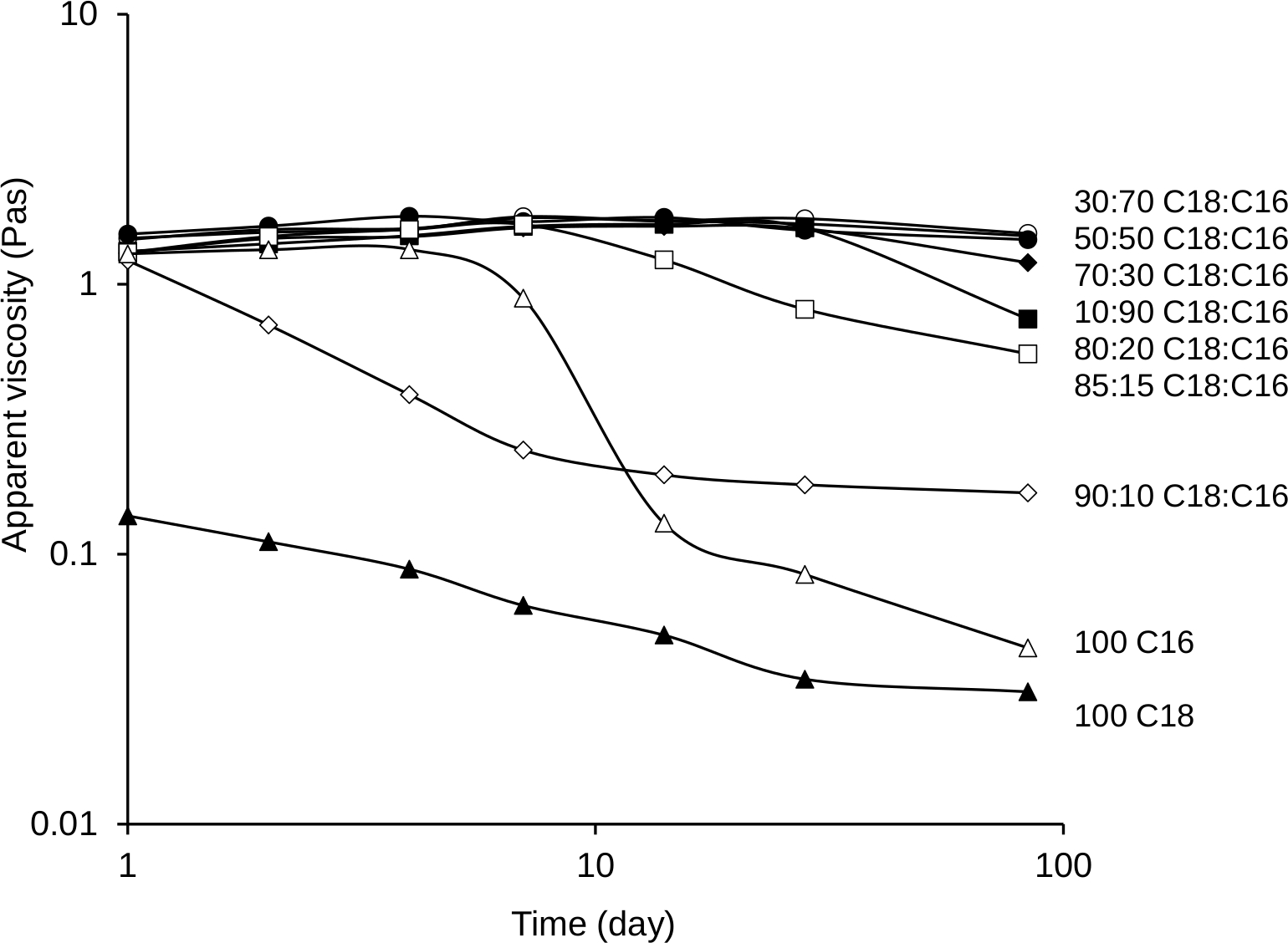} \\[\abovecaptionskip]
	\caption{Variation of apparent viscosity of all ternary systems (ratio C18:C16) vs. storage time.}
	\label{fig4}
\end{figure}

Flow curves for all the ternary systems were in the form of anti-clockwise hysteresis loops signifying their shear thinning behavior 
i.e. any shear stress-shear rate curve (plotted on double logarithmic coordinates) could be approximately fitted by a straight line over the shear rate sweep range under study; according to the Power Law model 
$\eta = K\dot{\gamma}^{n-1}$
(where $\eta$ is the viscosity, $K$ is the consistency coefficient, $\dot{\gamma}$ is the shear rate, and $n$ is the Power Law Index for a shear thinning fluid with $0<n<1$).
Continuous shear data were categorized into 3 groups pursuant to the fatty alcohol mixing ratios used. In group 1 (containing from 30\% to 70\% C18), the semisolid structure developed considerably when being aged for the first week, which was succeeded by a small consistent decline (e.g. \fref{fig2}a, \ref{fig2}b). In group 2 (containing 90\% C16 or 80--85\% C18) an consistent increase was initially noted, but a rapid reduction in the apparent viscosities was seen on further storage (e.g. \fref{fig3}a, \ref{fig3}b). On the other hand, a significant structure breakdown with storage time was rheologically observed for group 3 (containing pure C16, C18, or 90:10 C18:C16). The inconsistency of apparent viscosity $\eta_{app}$, derived at the apex of each loop (100s$^{-1}$), with storage time is plotted in \fref{fig4}.

\begin{figure}[!ht]
	\centering
	\includegraphics[width=.7\linewidth]{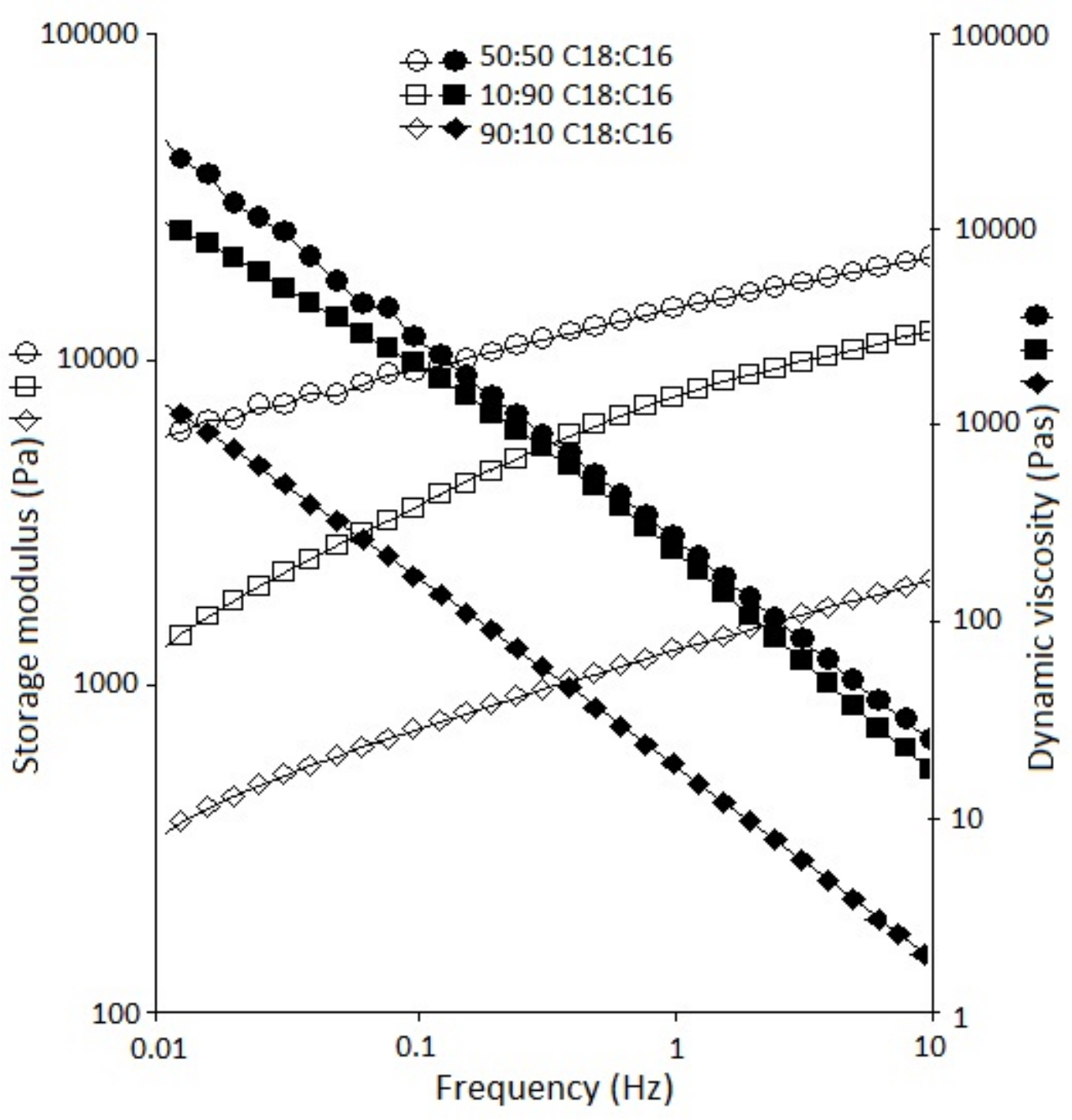} \\[\abovecaptionskip]
	\caption{Oscillation data (storage modulus $G'$ and dynamic viscosity $\eta'$)of some typical C18-C16 ternary systems after 3-month storage.}
	\label{fig5}
\end{figure}

\Fref{fig5} shows oscillatory data of representative ternary systems indicating that viscoelastic parameters ($G'$ and $\eta'$) of 50:50 C18-C16 system were the greatest over 3-month storage at 25\textsuperscript{o}C. Both continuous shear and viscoelastic data were in good agreement e.g. both apparent viscosity $\eta_{app}$ (accessed by continuous shear) and dynamic viscosity $\eta'$ (obtained by oscillation frequency sweep) values were in the same descending order for C18-C16 systems: 50:50 $>$ 10:90 $>$ 90:10.

\subsection{DSC data}
\begin{figure}[!ht]
	\centering
	\includegraphics[width=.6\linewidth]{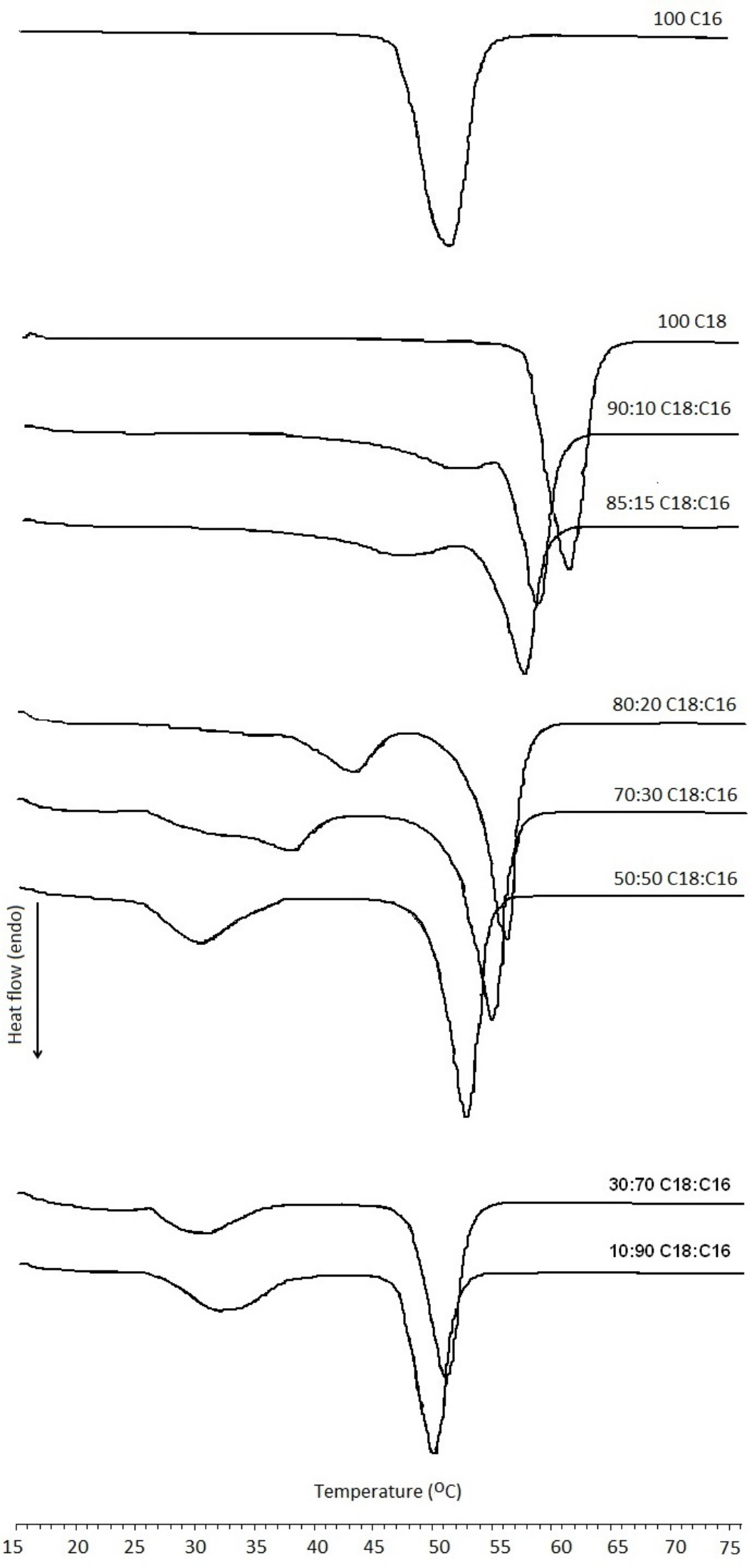} \\[\abovecaptionskip]
	\caption{Typical DSC diagrams of fatty alcohol admixtures.}
	\label{fig6}
\end{figure}

\begin{figure}[!ht]
	\centering
	\includegraphics[width=.67\linewidth]{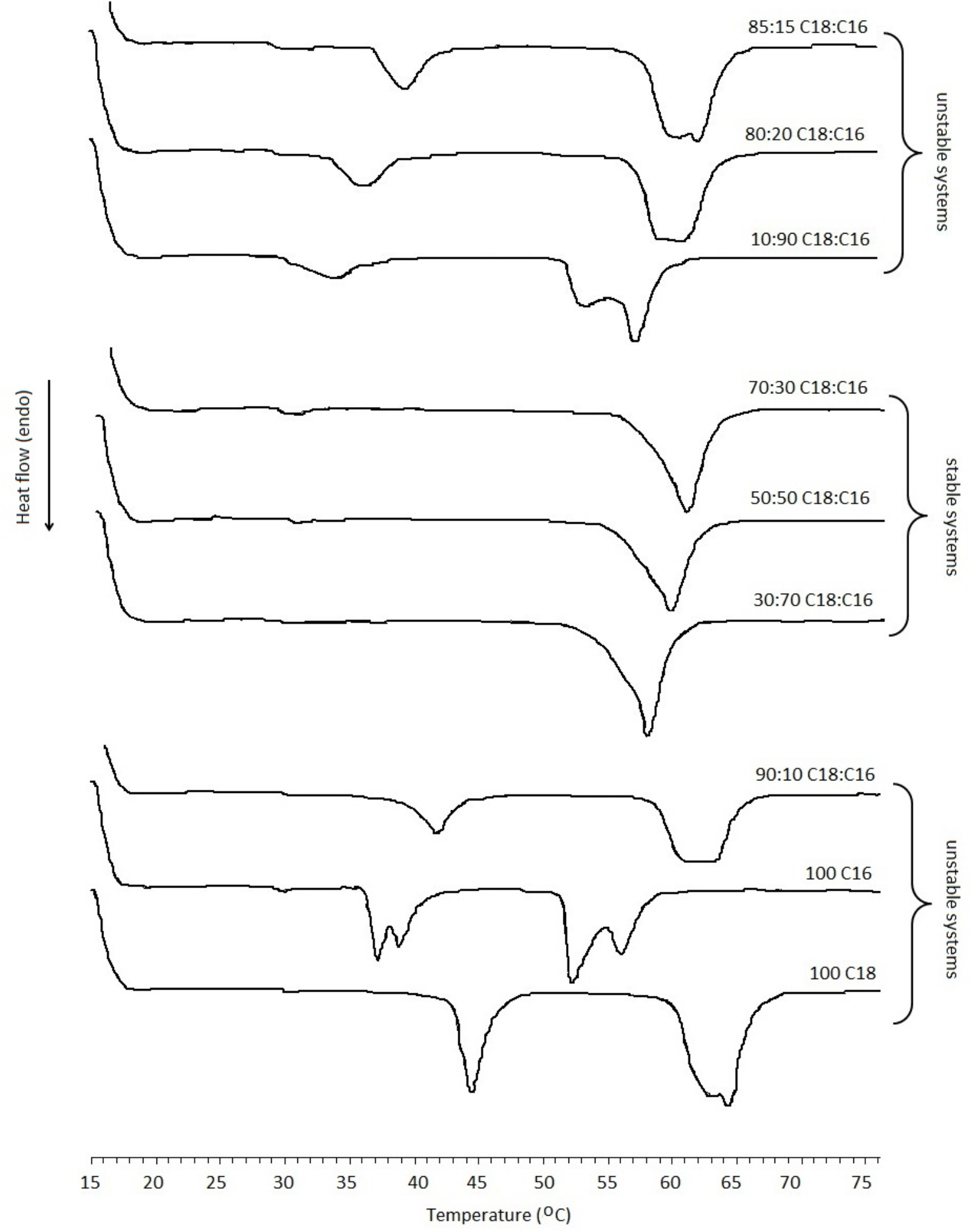} \\[\abovecaptionskip]
	\caption{Typical DSC diagrams of 3-month old ternary systems containing
		Polysorbate 60 and different ratio of C18:C16 fatty alcohols.}
	\label{fig7}
\end{figure}

\begin{figure}[!ht]
	\centering
	\includegraphics[width=.6\linewidth]{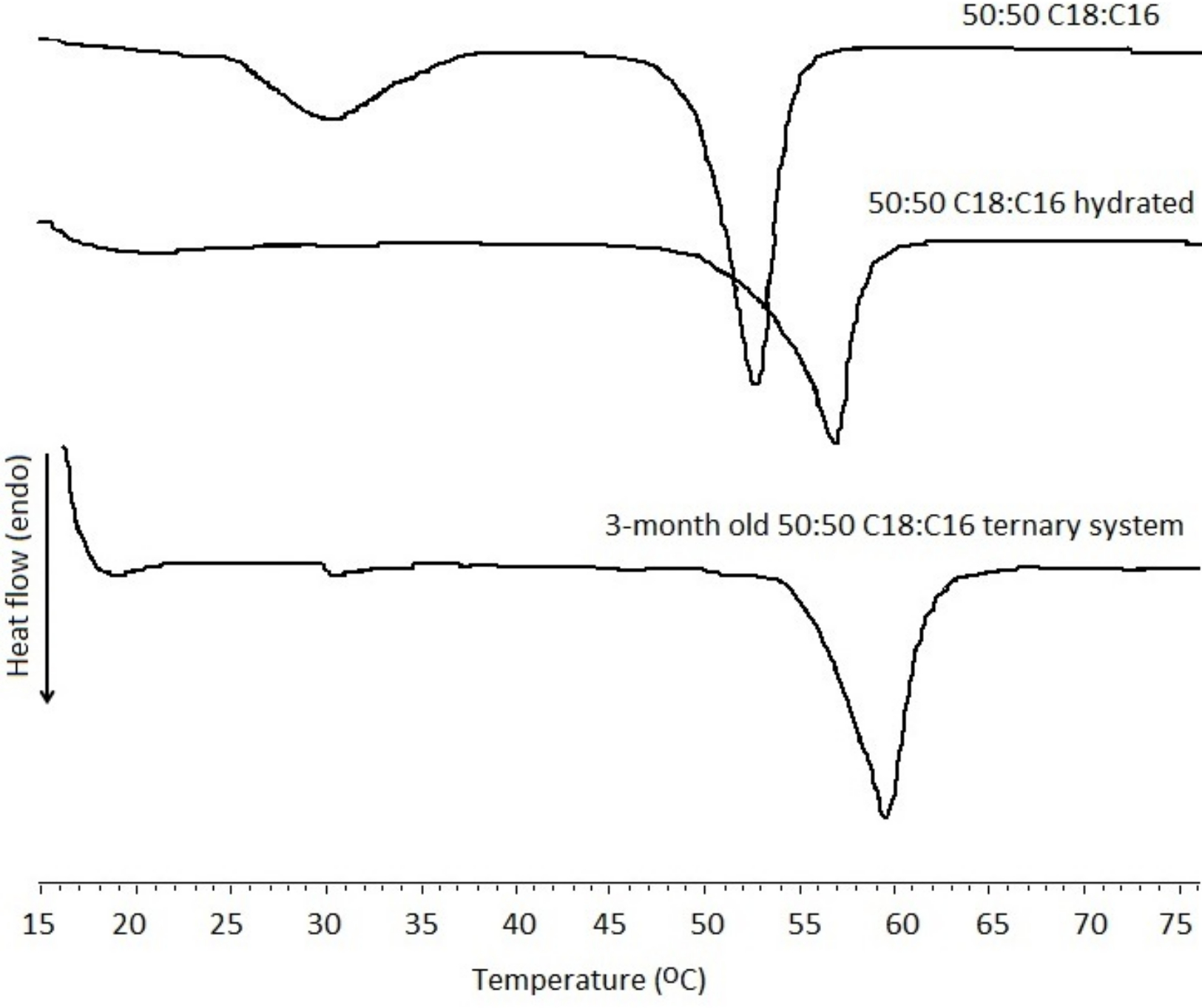} \\[\abovecaptionskip]
	\caption{Typical DSC diagrams of representative fatty alcohol admixture 50:50 C18:C16, their
		corresponding hydrated admixture and 3-month ternary system.}
	\label{fig8}
\end{figure}

In order to possibly correlate phase transitions observed in the ternary systems to the polymorphic transition and melting of fatty alcohols used, the DSC experiments of pure fatty alcohol homologues (C18, C16) and their mixtures of different ratios were carried out in parallel with corresponding ternary systems (figures \ref{fig6}, \ref{fig7}, \tref{table1}). Additionally, the effect of water on the phase transition of fatty alcohol admixture was also thermally studied (e.g. hydrated 50:50 C18:C16 shown in \fref{fig8}). 

\begin{table}[]
	\caption{DSC Onset temperatures of admixtures of fatty alcohols and their corresponding ternary systems.}
	\begin{tabular}{rcccccccccccccccccc}
		\br
		& \multicolumn{18}{c}{DSC onset temperatures of C18:C16 systems (\textsuperscript{o}C)} \\ \cline{2-19}
		& \multicolumn{6}{c}{Group 1}  &  \multicolumn{6}{c}{Group 2}  &  \multicolumn{6}{c}{Group 3} \\ \cline{2-19}
		& \multicolumn{2}{c}{85:15} & \multicolumn{2}{c}{80:20} & \multicolumn{2}{c}{10:90} & \multicolumn{2}{c}{70:30} & \multicolumn{2}{c}{50:50} & \multicolumn{2}{c}{30:70} &  \multicolumn{2}{c}{90:10} & \multicolumn{2}{c}{100C16} & \multicolumn{2}{c}{100C18} \\ \mr
		Peaks *          & 1  & 2  & 1  & 2  & 1  & 2  & 1  & 2  & 1  & 2  & 1  & 2  & 1  & 2  & 1  & 2  & 1  & 2  \\ \mr
		\multicolumn{1}{c}{\begin{tabular}[l]{@{}c@{}}Fatty\\ alcohols\end{tabular}}
		& 32 & 51 & 34 & 45 & 25 & 42 & 25 & 45 & 25 & 45 & 25 & 45 & 45 & 54 & 47 & -  & 55 & -  \\ 
		\multicolumn{1}{c}{\begin{tabular}[c]{@{}c@{}}Ternary\\ systems\end{tabular}}
		& 38 & 56 & 34 & 56 & 30 & 51 & 53 & -  & 53 & -  & 51 & -  & 38 & 57 & 36 & 50 & 42 & 58 \\ \br
	\end{tabular}
	\label{table1}
	\begin{indented}
	\item[] * 1: first peak, 2: second peak
	\end{indented}
\end{table}

\begin{figure}[!ht]
	\centering
	\includegraphics[width=.67\linewidth]{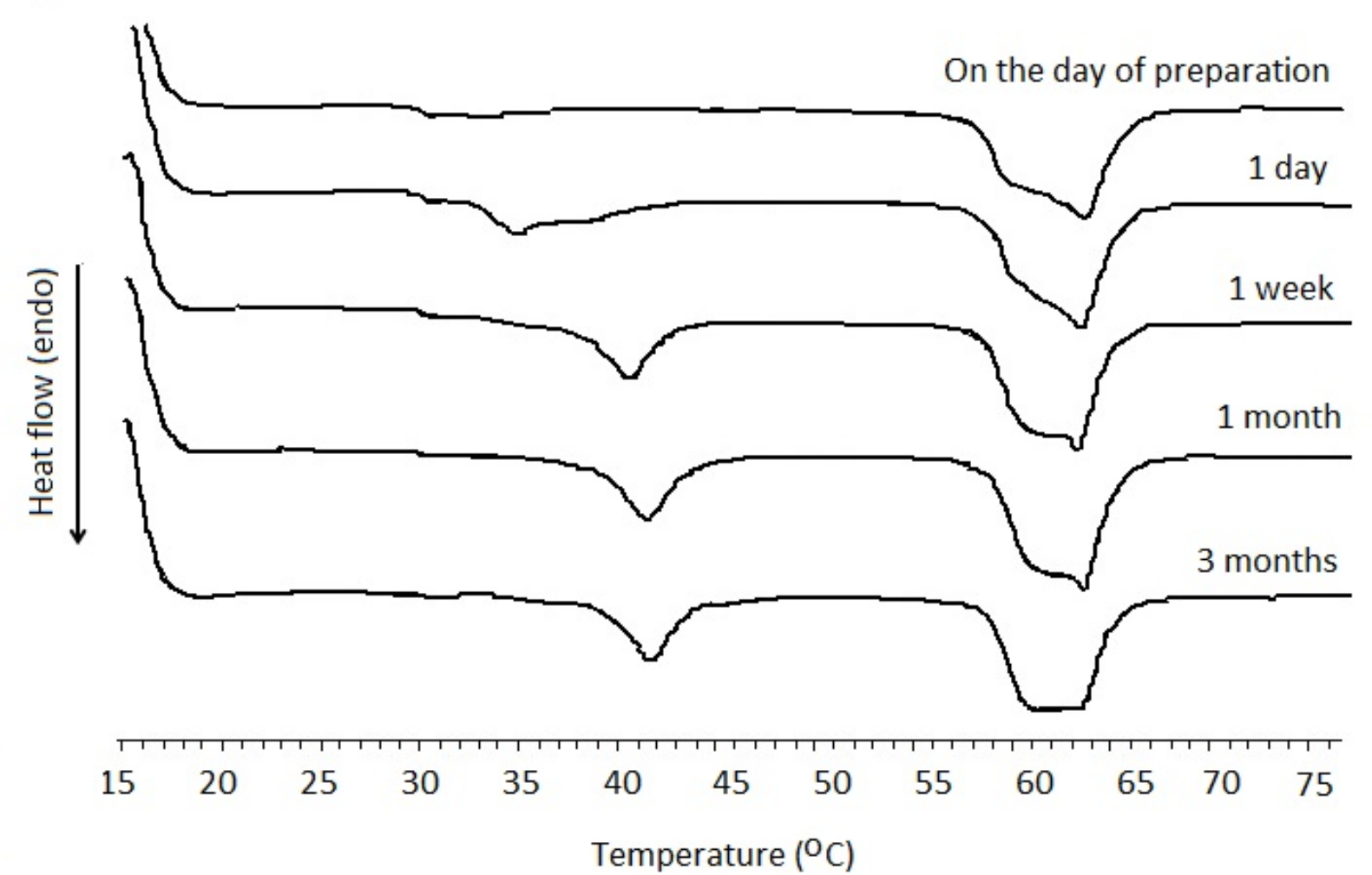} \\[\abovecaptionskip]
	\caption{Typical DSC diagrams of 90:10 C18-C16 system during 3-month storage.}
	\label{fig9}
\end{figure}

Apart from pure fatty alcohol homologue, each admixture shows two broad asymmetrical peaks in the range of ca. 25--63\textsuperscript{o}C depending on the C18:C16 ratios. In each stable ternary system (e.g. 50:50 C18:C16), there was only a broad asymmetrical endothermic peak starting around 53\textsuperscript{o}C and peaking at ca. 60\textsuperscript{o}C. On the contrary, the unstable ternary systems generated several endothermic transitions as they weakened and decomposed. For instance, in 90\% C18 system the spectrum acquired (when it was semisolid) was similar to that acquired for stable systems i.e. only a broad asymmetrical endotherm peaking at ca. 63\textsuperscript{o}C. As this system broke down, this transition broadened somewhat and another endothermic peak, which increased in intensity on storage, developed at ca. 40\textsuperscript{o}C (\fref{fig9}).

\subsection{Simulation of fatty alcohol--polysorbate 60--water ternary systems}
\begin{figure}[!ht]
	\centering
	\includegraphics[width=1\linewidth]{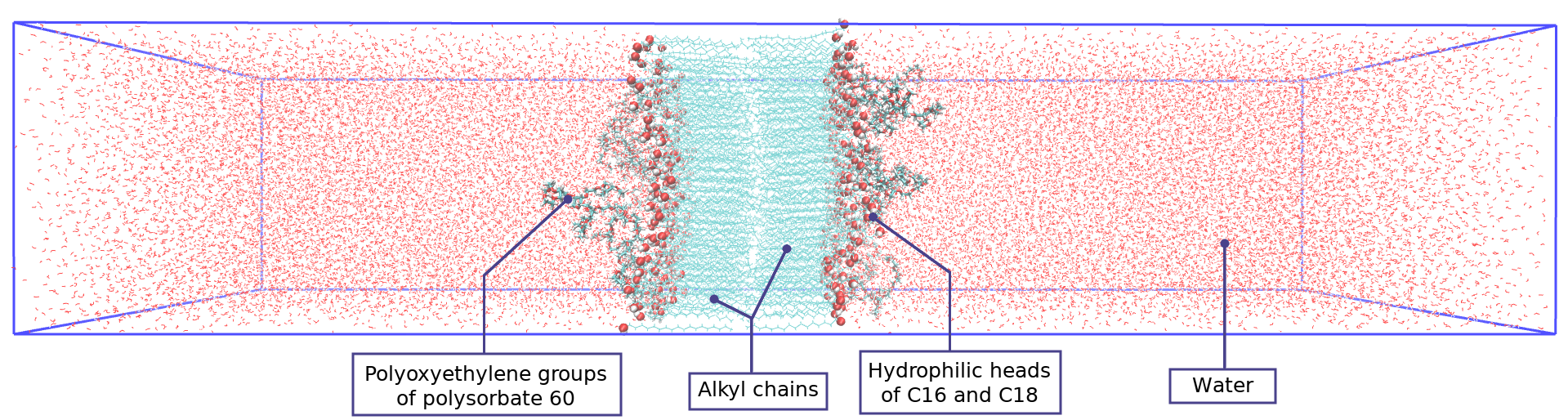} \\[\abovecaptionskip]
	\caption{A snapshot of the 50:50 C18:C16 system after being simulated for 750ns showing the typical arrangement of the bilayer membrane fluctuating in water.}
	\label{fig10}
\end{figure}

To investigate the effect of mixing ratios of C18-C16 alcohols, computational molecular dynamics simulations were carried out. \Fref{fig10} is a snapshot of the 50:50 C18:C16 system during the production run after 750ns showing typical components and structure of the simulated system. The thickness of water between two adjacent membranes is about 28nm within the same range of typical experimental systems. All the simulations were performed for 1000ns. Then, from all the configurations of systems, the area per lipid and deuterium order parameter (SCD) are calculated for C16, C18, and polysorbate 60. \Tref{table3and4} shows the SCD of C18 and C16 molecules (all error bars are less than 0.003). The value of SCD varies depending on the position of C-atoms of the hydrocarbon chain implanting in the membrane, with the lowest values (less order) at the two extremes of the chain and the highest values (more order) in the middle. \Tref{table6} shows the area per lipid of each type of molecules, C18, C16, polysorbate 60, and the average area per lipid of the whole membrane. As the mixing ratio changes, the area per lipid also changes linearly around 19.4$\AA{}^2$, but not significantly. However, the magnitude of fluctuations in the area per lipid changes considerably. 

\Table{\label{table3and4} Deuterium order parameter of C18 (column 2 to 6) and C16 (column 7 to 11) alcohol molecules. The C-atoms are numbered from the farthest to the nearest C-atom to the head (OH group) of alcohol molecules.}
\br
C-atom & 10:90 & 30:70 & 50:50 & 70:30 & 90:10 & 10:90 & 30:70 & 50:50 & 70:30 & 90:10\\ \mr
2      & 0.324 & 0.331 & 0.351 & 0.371 & 0.395 & 0.403 & 0.408 & 0.418 & 0.426 & 0.436\\
3      & 0.381 & 0.386 & 0.401 & 0.417 & 0.435 & 0.438 & 0.441 & 0.447 & 0.453 & 0.458\\
4      & 0.430 & 0.432 & 0.441 & 0.450 & 0.459 & 0.459 & 0.460 & 0.464 & 0.465 & 0.469\\
5      & 0.454 & 0.455 & 0.459 & 0.463 & 0.468 & 0.465 & 0.466 & 0.468 & 0.469 & 0.470\\
6      & 0.462 & 0.463 & 0.466 & 0.468 & 0.470 & 0.467 & 0.468 & 0.469 & 0.471 & 0.472\\
7      & 0.466 & 0.467 & 0.469 & 0.470 & 0.472 & 0.468 & 0.470 & 0.471 & 0.471 & 0.473\\
8      & 0.467 & 0.468 & 0.470 & 0.471 & 0.473 & 0.469 & 0.470 & 0.471 & 0.472 & 0.474\\
9      & 0.468 & 0.468 & 0.471 & 0.471 & 0.473 & 0.469 & 0.470 & 0.471 & 0.472 & 0.473\\
10     & 0.469 & 0.469 & 0.471 & 0.472 & 0.473 & 0.468 & 0.469 & 0.471 & 0.472 & 0.473\\
11     & 0.467 & 0.469 & 0.471 & 0.472 & 0.473 & 0.467 & 0.468 & 0.470 & 0.471 & 0.472\\
12     & 0.467 & 0.468 & 0.470 & 0.471 & 0.472 & 0.465 & 0.467 & 0.468 & 0.469 & 0.471\\
13     & 0.466 & 0.466 & 0.468 & 0.470 & 0.471 & 0.461 & 0.462 & 0.464 & 0.466 & 0.469\\
14     & 0.461 & 0.463 & 0.465 & 0.467 & 0.469 & 0.448 & 0.449 & 0.453 & 0.456 & 0.461\\
15     & 0.452 & 0.455 & 0.458 & 0.461 & 0.465 & 0.411 & 0.415 & 0.420 & 0.427 & 0.436\\
16     & 0.424 & 0.430 & 0.439 & 0.443 & 0.451 & 0.294 & 0.302 & 0.311 & 0.324 & 0.340\\
17     & 0.366 & 0.377 & 0.390 & 0.400 & 0.414 &&&&&\\
18     & 0.231 & 0.249 & 0.261 & 0.271 & 0.288 &&&&&\\ \br
\endTable


\Table{\label{table6} The area per lipid (in $\AA{}^2$) for each type of molecules in all simulated systems.}
\br
System & C16          & C18          & Polysorbate 60 & Total        \\ \mr
10:90  & 19.48 $\pm$ 0.13 & 19.46 $\pm$ 0.43 & 19.19 $\pm$ 0.56   & 19.46 $\pm$ 0.10 \\
30:70  & 19.44 $\pm$ 0.13 & 19.42 $\pm$ 0.23 & 19.16 $\pm$ 0.53   & 19.41 $\pm$ 0.10 \\
50:50  & 19.41 $\pm$ 0.16 & 19.37 $\pm$ 0.17 & 19.16 $\pm$ 0.54   & 19.38 $\pm$ 0.10 \\
70:30  & 19.36 $\pm$ 0.20 & 19.32 $\pm$ 0.13 & 19.04 $\pm$ 0.53   & 19.32 $\pm$ 0.09 \\
90:10  & 19.31 $\pm$ 0.38 & 19.25 $\pm$ 0.13 & 19.00 $\pm$ 0.51   & 19.24 $\pm$ 0.10 \\ \br
\endTable

\section{Discussion}\label{discussion}
Generally, C16, C18 fatty alcohols have at least three polymorphic forms i.e. $\alpha$, $\beta$, $\gamma$. At melting temperature, the separation of $\alpha$-polymorph occurs. This form is durable over a narrow range of temperatures, with hydrocarbon chains hexagonally packed and fully extended in the transformation as well as rotational motion about the long molecular axis. At lower temperatures, the $\beta$-form (with non-rotating hydrocarbon chains) co-exists with the tilted $\gamma$-form although the $\beta$-form is normally in excess \cite{Tanaka1958}. On heating, $\beta$-$\gamma$ transition scarcely occurs while $\beta$-$\alpha$ transition does happen at a temperature somewhat higher than $\gamma$-$\alpha$ transition point \cite{Fukushima1977}. Of particular interest is the $\alpha$-form, for it is able to form spontaneously swollen $\alpha$-crystalline gel phase when the dispersion of fatty alcohol in water takes place in the occurrence of a small quantity of surfactant \cite{Eccleston1997}. The DSC data for pure fatty alcohol and their mixture obtained in this study reasonably agree well with literature information \cite{Eccleston1985, Fukushima1976, Vringer1986}. The broad endotherm peaking at 53\textsuperscript{o}C (range 46--58\textsuperscript{o}C) for pure C16 or at 62\textsuperscript{o}C (range 57--67\textsuperscript{o}C) for pure C18 combines the $\beta$, $\gamma$-$\alpha$ transition and melting of the $\alpha$-form. With a C16-C18 admixture, lowering transition temperatures and extending polymorphic ranges are observed \cite{Fukushima1977, Stewart1960}. In \Fref{fig6}, the DSC diagrams of C16-C18 mixtures reveal the broad endotherms over lower temperature range (representing $\beta$, $\gamma$-$\alpha$ transition that is even further depressed in the presence of water (\fref{fig8})); and the sharp endotherms peaking at higher temperature due to the melting of the $\alpha$-form.

In our study, ternary systems were prepared by homogenizing 70\textsuperscript{o}C excipients and then cooling whilst blending to room temperature (ca. 25\textsuperscript{o}C). In order to understand thoroughly, the mechanism and kinetics involved in the formation and initial consistent increase of all ternary systems, the gel network theory proposed by Eccleston and Barry is referred to \cite{Eccleston1997}. This gel network theory has been extensively used for explanation in literature so far (e.g. \cite{Ballman2008, Konya2003, Snezana2005})

At hydrocarbon chain-melting temperature, smectic liquid crystals are formed by the interaction of fatty alcohol with polysorbate 60 molecules. As the system cools down, a complex viscoelastic network is formed by the precipitation of the liquid crystals. Therefore, after the heating and cooling cycle of preparation, the ternary system may consist of no less than three phases: (i) crystalline hydrate of fatty amphiphile (microscopically revealed as anisotropic clumps) in balance with (ii) swollen lamellar $\alpha$-crystalline gel phase, microscopically revealed as Maltese crosses which can be also clumped together \cite{Eccleston1986} and (iii) free bulk water. The results of DSC experiments (\fref{fig7} and \tref{table1}) show that the existence of ternary systems can be available in several forms over a certain range of temperatures. When being stored at 25\textsuperscript{o}C, the stable form depends upon the C16:C18 ratio in each system.

Fundamentally, the structural changes indicated by rheological data (figures \ref{fig2}--\ref{fig5}) can be thoroughly explained as follows. The initial consistent increases are due to the increased solubility of polyoxyethylene (POE) groups in polysorbate 60 molecules on storage that allows additional lamellar gel phase to form. A slight reduction in consistency for 30--70\% C18 systems is probably owing to the re-arrangement of polysorbate 60 molecules among the viscoelastic gel network. In these stable systems, the mixture of C16-C18 alcohols creates sufficient disorder and inhibits crystallization (i.e. no evidence of crystals existed in these systems as representatively shown in \fref{fig1}a) so that the viscoelastic gel network was stable over 3-month storage at 25\textsuperscript{o}C. The broad endotherm peaking at ca. 63\textsuperscript{o}C in these systems represents the transition of hydrocarbon chains in the lipophilic bilayer from the $\alpha$-form to the liquid state. 

In contrast, in unstable systems a polymorphic conversion (from the $\alpha$-form to the more stable $\beta$-, $\gamma$-forms) appears, which was illustrated by an increase in intensity of the low-temperature endotherm starting at ca. 32\textsuperscript{o}C and peaking at ca. 40\textsuperscript{o}C on storage (e.g. 90:10 C18-C16 system shown in \fref{fig9}). Additionally, the high broad endotherm was resolved as a doublet (e.g. 10:90 C18-C16 system shown in \fref{fig7}) with the shoulder at the low-temperature side of this peak presumably connected with the region of hydrated fatty alcohol, in which no polysorbate 60 is incorporated \cite{Vringer1986}. Furthermore, the $\beta$, $\gamma$-polymorphs are crystallized, evidenced as tiny needles or diamond-shaped crystals in \fref{fig1}b. Thus, the reduction of the amount of interlamellar water gradually progresses and the viscoelastic gel phase collapses.

Data obtained in our study are in good agreement with previous results provided by Eccleston \cite{Eccleston1976, Eccleston1982, Eccleston1985, Eccleston1986} and other authors \cite{Barry1970, Talman1970}. Nevertheless, our results are somehow different from some observations, in which stable oil-in-water emulsions containing C16-C18 alcohols could be obtained with the ratios of C18 up to $\geq$ 90\% \cite{Barry1970} or C16 up to $\geq$ 90\% and Fukushima and co-workers \cite{Fukushima1976} showed that pure C18 was shown to give more stable oil-in-water emulsion than pure C16; while in our unstable systems, the greater amount of C18 the formulation contained the faster the rate of structure breakdown was. Perhaps, the striking differences in bulky nature of non-ionic surfactants used in our work (i.e. POE 20 sorbitan monostearate) and others’ surfactants i.e. single POE chain non-ionic, POE 15 oleyl ether, sodium dodecyl sulfate have essentially affected the instability of the gel network prepared with them \cite{Barry1970, Fukushima1976}. From the applicability point of view, it is worth mentioning that C16 and C18 alcohols are the main components of cetostearyl alcohol (as specified in British Pharmacopoeia 2016) \cite{BP2016}, which is a mixed homologue of C10, C12, C14, C16, C18, and C20. However, the emulsions containing the same ratios of long-chain fatty alcohols found in a cetostearyl alcohol batch commercially available were still slightly less stable than emulsion prepared with cetostearyl alcohol. It was possibly because of trace amounts of compounds unidentified by gas chromatography in the commercial cetostearyl alcohol \cite{Barry1970}.

From the molecular dynamics simulations, some observations on the effect of mixing ratios of C18-C16 alcohols are revealed as follows. 

The first observation is that the SCD of C-atoms at both the two ends of C18 molecules in the membrane decreases as the percentage of C16 increases (\tref{table3and4}). SCD is a measure of the flexibility of the hydrocarbon chain of a molecule. The closer SCD gets to 0.5, the more rigid the structure of a molecule is. For a very flexible structure, SCD tends to decrease to zero. \Tref{table3and4} shows that both the head and tail parts of C18 molecules become more flexible when C16 molecules are present in the membrane. This observation supports well the experimental hypothesis in that the stability of the membrane is due to a higher configurational entropy of the system due to the alkyl chain length mismatch of cetearyl alcohol \cite{Iwata2017}. Our results provide an additional insight i.e. the tail of C16 molecules becomes stiffer and more ordered (SCD increases) as the mixing percentage of C18 increases (\tref{table3and4}). This allows for even more spatial flexibility of the longer alcohol chains. 

The second observation is that the area per lipid has the fluctuations of C16 and C18 increased as the mixing ratio of C16 and C18 is gradually equal to 1 (\tref{table6}). In other words, the 50:50 C18:C16 system has the maximum fluctuations of C16 and C18 in comparison with those of other systems. In detail, the fluctuation of C16 in the 50:50 C18:C16 system is 0.16, which is about 20\% higher than that of C16 in the 10:90 C18:C16 system. The fluctuations of C18 follow the same pattern i.e. the fluctuation of C18 in the 50:50 C18:C16 system is about 30\% higher than that of the 90:10 C18:C16 system.

The above-mentioned observations strongly support that at the mixing ratio 50:50 of C18 and C16, the membrane is most flexible, with a higher configurational entropy in comparison with others, namely 10:90, 30:70, 70:30, and 90:10. This finding is consistent with the experimental hypothesis and provides additional molecular details to this hypothesis. It also agrees well with the observed stability of the 50:50 C18:C16 system shown in many experiment measurements presented above \cite{Iwata2017, Snyder1983}.

\section{Conclusion}\label{conclusion}
The consistency and stability of all ternary systems prepared with polysorbate 60 and combinations of various C18:C16 fatty alcohol ratios have been explained by the gel network theory describing a relationship between the structure and physico-chemical properties of ternary systems to the swelling properties of $\alpha$-swollen lamellar gel phase. Our experimental data show that the consistency of ternary systems containing from 30\% to 70\% C18 did not change significantly over 3-month storage at 25\textsuperscript{o}C, whereas some structure breakdown was seen with the remaining systems, especially with C18-C16 combinations in which one fatty alcohol was in excess of the other ($\geq$ 90\%). The 50:50 C18:C16 system shows the highest consistency with the highest storage modulus and dynamic viscosity after 3-month storage. Instabilities in these systems may be related to the polymorphic transition from $\alpha$ to $\beta$, $\gamma$ and crystallization of fatty alcohols used. These findings correlate well with other studies and strongly suggest the use of a large number of various fatty amphiphiles and surfactants within the same formulation to counteract the effect of batch variation of emulsifier components.

Computational simulations using an atomistic model of explicit solvent were performed at different mixing ratios of the fatty alcohols to investigate the systems at molecular details. Various parameters of ordering in the systems were analyzed. It lends strong support to the experimental hypothesis that the higher stability of mixed systems is by virtue of an increase in configurational entropy related to a mismatch in alkyl chain length. Additional details on the mechanism are obtained from simulations. The alcohol with shorter alkyl chain actually becomes more rigid. This helps to further increase the configuration space for the alcohol.

Our combined experimental and computational investigations agree with each other and provide a comprehensive, complementary understanding of these systems at macroscopic and molecular levels. This modern approach can be profitably exploited for developing pharmaceutical cream formulation based on the quality by design approach.

\section*{Acknowledgment}
This research is supported by Vietnam National Foundation for Science and Technology Development (NAFOSTED) under grant number 104.99-2016.39.

\section*{References}
\bibliographystyle{iopart-num}
\bibliography{tween60}

\end{document}